\documentclass[12pt,preprintnumbers,showpacs,amsmath,amssymb,prd,floatfix,superscriptaddress,nofootinbib]{revtex4}
\usepackage{graphicx}
\usepackage{epsfig}
\usepackage{bm}
\usepackage{amsfonts}
\begin{document}
\title{A Slowly Rotating Black Hole in
Horava-Lifshitz Gravity and a $\textbf{3+1}$ Dimensional Topological
Black Hole: Motion of Particles and BSW Mechanism}
\author{Ibrar Hussain}
\email{ibrar.hussain@seecs.nust.edu.pk}
\affiliation{School of Electrical Engineering and Computer Science\\
National University of Sciences and Technology\\
H-12, Islamabad, Pakistan.}
\author{Mubasher Jamil}
\email{mjamil@sns.nust.edu.pk}
\affiliation{School of Natural Sciences, National University of Sciences and Technology,\\
H-12, Islamabad, Pakistan.}

\author{Bushra Majeed}
\email{bushra_majeed18@yahoo.com}
\affiliation{School of Natural Sciences, National University of Sciences and Technology,\\
H-12, Islamabad, Pakistan.}

\begin{abstract}
The motion of a neutral particle in the vicinity of a slowly
rotating black hole in the Horava-Lifshitz theory of gravity and 3+1
dimensional topological Lifshitz black hole is investigated.
Geodesics for radial motion of the particles are also plotted. Some
different cases of the orbital motion of the particle are discussed
where maximum and minimum values of the effective potential are
calculated. Further the Ba\~{n}ados, Silk and West (BSW) mechanism
is studied for these black holes. It is shown that the
centre-of-mass energy (CME) of two colliding uncharged particles at
the horizon of these black holes remains finite. Thus the BSW effect
cannot be seen in these cases.

\text {Key words}:  Motion of particles; Black holes;
Horava-Lifshitz gravity, 3+1 dimensional topological Lifshitz black
hole; BSW mechanism.
\end{abstract}

 \maketitle
\newpage

\section{Introduction}

To combine the Einstein's theory of relativity with quantum
mechanics or to obtain a theory of quantum gravity (QG) is an
unsolved problem in physics. In this regard several attempts have
been made (for detail see \cite{QG}). In 2009 Peter Horava proposed
a candidate theory for QG in 4 spacetime dimensions which is based
on the explicit violation of local Lorentz invariance\cite{PH}. This
is a power counting renormalizable, higher order gravity model and
reduces to General Relativity (GR) at large distances. It may
provide a candidate of ultraviolet completion of GR \cite{PH}.

In the past few years some static and slowly rotating black hole
spacetimes have been proposed in Horava-Lifshitz (HL) gravity
\cite{KS, LKM, Ci, BS1,BS2, BJS, LMP}. In this regard Lu-Mei-Pope
have studied a spherically symmetric black hole with some dynamical
parameter $\lambda$ \cite{LMP}. Cai et. al have obtained a
topological black hole solution\cite{Ci}. Another black hole
solution which is asymptotically flat has been found by Kehagias and
Sfetsos by taking $\lambda=1$ \cite{KS}. Besides, some slowly
rotating black hole solutions in HL gravity have been investigated
in literature \cite{LKM, BS1,BS2, BJS}.

Recently researchers have given a great attention to the Lifshitz
spacetimes for finding the gravity duals of Lifshitz fixed points
due to the AdS/CFT correspondence for condensed matter physics
\cite{mann1}. From the perspective of quantum field theory, there
are many scale invariant theories for studying such critical points.
Lifshitz field theory is a toy model which exhibits this scale
invariance. Such theories exhibit the anisotropic scale invariance
expressed as $t\rightarrow\lambda ^z t$, and $x \rightarrow\lambda
x$,  with $z\neq1$, where $z$ is the relative scale dimension of
time and space. The Lifshitz spacetimes are given by
\begin{equation}\label{topo1}
ds^2 =  \frac{r^{2z}}{\ell^{2z}} dt^2 -
\frac{\ell^2}{ r^2} dr^2 - \frac{r^2}{ \ell^2}
d{\vec{x}}^2,
\end{equation}
where $\vec{x}$ is a $D - 2$ dimensional spatial vector, $D$ denotes
the spacetime dimension and $\ell$ represents the length scale in
the geometry. As mentioned, this spacetime is interesting due to its
invariance under anisotropic scale transformation and represents the
gravitational dual of strange metals \cite{mann2}. When $z = 1$,
then (\ref{topo1}) is the usual anti-de Sitter metric in
Poincar\'{e} coordinates. The metric of the Lifshitz black hole
is asymptotically similar to (\ref{topo1}). Topological Lifshitz
black hole with critical exponent $z=2$ in $3+1$ dimensions was
proposed by Mann \cite{2009}. Particles motion on topological black
hole in $3+1$ dimensions has been studied in \cite{topological}.

Particle dynamics around black holes is a topic of interest for
physicist for the last few decades. In this direction several papers have
been appeared in the literature. Particle motion in the vicinity of
a magnetized Schwarzschild black hole was studied in \cite{Zah}.
Particle dynamics in the Riessner-Nordstrom, Kerr and Kerr-Newman
black holes was investigated by Pugliese et. al.
\cite{RR1,RR2,RR3,RR4}. An analysis of motion of particles in higher
dimensional black holes has appeared in \cite{VS}. Vasudevan et. al.
have investigated particle motion in Myers-Perry black hole in all
dimensions \cite{Vs}. The motion of particles around near-extreme
Braneworld Kerr black holes has analyzed in \cite{Br}. In a regular
black hole spacetime, motion of particles has discussed by Garcia
et. al \cite{Ga}. Particles motion in the background of a black hole
in Branward geometry has investigated in \cite{AA1}. Enolskii et.
al. considered motion of particles in the Kehagias and Sfetsos black
hole \cite{Eno}. They have shown that neither massless nor massive
particles with non-vanishing angular momentum can reach the
singularity. The dynamics of particles around Kehagias and Sfetsos
black hole immersed in an external magnetic field has studied in
\cite{AA2}. In \cite{AA3} motion of charged particles in a rotating
black hole immersed in a uniform magnetic field has considered. The
influence of quintessence on the motion of particles in the vicinity
of Schwarzschild black hole has discussed by Fernando \cite{SF} and
Li and Zhang \cite{EZ}. The tunneling radiation characteristic of charged
particles from the Reissner-Nordström-anti de Sitter black hole has
studied in \cite{DQS}.

Ba\~{n}ados, Silk and West (BSW) proposed that rotating black holes
may act as particle accelerators \cite{BSW}. In the vicinity of
extremal Kerr black hole, they have found that infinite centre-of-mass
energy (CME) can be achieved during the collision of particles. The
BSW effect was then studied for different black hole spacetimes
\cite{JS, OB, GP, SYH, La,OB1, SYC, SYH1, YSY, AW, MP, ZG1, PRY,
ZG2, BCG, IH1, IH2, IH3, MS1, MS2}. For a brief review one can see
\cite{IH1} and references therein. BSW effect for particles in the
vicinity of a rotating black hole in HL gravity has studied by
Abdujabbarov and Ahmedov in \cite{AA2}. They have found some limitation
on the CME of the accelerating particles. For another slowly
rotating black hole in HL gravity the BSW mechanism has investigated
by Sadeghi and Pourhassan \cite{JB1}. They have obtained an infinite CME
for colliding particles arbitrarily close to the horizon of the
slowly rotating balck hole. Consequently in this paper we consider a
slowly rotating black hole solution in HL theory of gravity given by
Barausse et. al \cite{BS2}, and the $3+1$ dimensional topological
Lifshitz black hole, to study motion of particles in the vicinity of
the black holes. We also investigate the BSW process for these black
hole to look at the CME of the colliding neutral particles.

The plan of the paper is as follows. In Section 2 we formulate the
equations of motion for a particle in the field of the slowly
rotating black hole in the HL gravity. In subsections geodesics of
the radially moving particle are plotted. We also analyze the
behavior of the effective potential to look at its maximum and
minimum values. Expression for CME of the colliding particles is
obtained. In Section 3 we study the motion of particles in the
vicinity of $3+1$ dimensional topological Lifshitz black hole.
Behavior of the effective potential and CME of the colliding
particles around this black hole are studied in the following
subsections. We summarize our discussion in the last Section.
Throughout this paper we use $G=c=1$.

\section{Motion of particles in the vicinity of slowly rotating black hole in
Horava-Lifshitz gravity}
The line element for a slowly rotating black hole in HL theory of
gravity is given by \cite{BS2}
\begin{equation}\label{smetric}
 ds^2=f(r)dt^2-\frac{B(r)^2}{f(r)}dr^2-r^2(d{\theta}^2+\sin^2{\theta}d\phi^2 )
+\epsilon r^2\sin^2\theta \Omega (r,\theta )dtd\phi + O(\epsilon^2),
\end{equation}
where $f(r)$ and $B(r)$ are given as a series expansion for the
spherically symmetric static asymptotically flat solution in terms
of the inverse radial coordinates $x=\frac{\alpha}{r}$ (for detail
one can see \cite{BJS}). For simplicity, we consider only those
terms
 which are linear in $x$ and neglect the higher order terms. Therefore
\begin{equation}
f(r)=1+\frac{\alpha}{r},\quad B(r)=1,
\end{equation}
 where $\alpha$ is an arbitrary
constant with dimensions of length. In (2) $\epsilon$ is known as the
book-keeping parameter of the expansion in the rotation. The motion
of a neutral particle in the vicinity of a black hole is determined
by the following Lagrangian
\begin{equation}\label{Lag}
 \mathcal{L}=\frac{1}{2}g_{\mu\nu}\dot{x^\mu}\dot{x^\nu},
\end{equation}
where $g_{\mu\nu}$ is the metric tensor and over-dot denotes the
derivative with respect to the geodetic parameter. Here $\mu,~~ \nu=
0,~ 1,~ 2, ~3$. Using (\ref{smetric}) in (\ref{Lag}), we get
\begin{equation}\label{1}
\mathcal{L}=\frac{1}{2}\Big[
f(r)\dot{t}^2-\frac{B(r)^2}{f(r)}\dot{r}^2-r^2\dot{\theta}^2-r^2\sin^2\dot\phi^2+\epsilon
r^2\sin^2\theta\Omega(r,\theta)\dot t\dot\phi +O(\epsilon^2) \Big].
\end{equation}
The time coordinate basis $\partial/\partial{t}$ and azimuthal
angular coordinate basis $\partial/\partial{\phi}$ are Killing
vectors of the metric (\ref{smetric}) we considered here, therefore
conserved quantities along the geodesics of the particles are the
energy $E$, of the particle and the angular momentum $L$, of
the particle respectively. The Euler
-Lagrange equations give the following relations upon integration
\begin{eqnarray}\label{2}
\tilde{\mathcal{E}}\equiv-\frac{E}{m}=\frac{\partial
\mathcal{L}}{\partial \dot t}&=&f(r)\dot
t+\frac{\epsilon}{2}(r^2\sin^2\theta) \Omega\dot \phi,\label{2}\\
\tilde{\mathcal{L}}\equiv -\frac{L}{m}=\frac{\partial
\mathcal{L}}{\partial \dot \phi}&=&
-r^2\sin^2\theta\dot\phi+\frac{\epsilon}{2}(r^2\sin^2\theta)\Omega\dot
t ,\label{3}
\end{eqnarray}
here $m$ is mass of particle and terms of order $O(\epsilon)$ are
retained while higher order terms are ignored in whole calculations.
With-out loss of generality, the motion of the particles can be
confined to the equatorial plane, $\theta=\pi/2$. For $\theta=\pi/2$
and eliminating $\dot t$ from (\ref{2}) and (\ref{3}), we get
\begin{equation}\label{4}
\dot\phi=\frac{L}{mr^2}-\epsilon\Big(  \frac{E\Omega}{2mf} \Big).
\end{equation}
 Further substituting (\ref{4}) in (\ref{2}), we get
\begin{equation}\label{5}
\dot t =\frac{-1}{mf}\Big(  E+\frac{\epsilon}{2}\Omega L \Big),
\end{equation}
To find $\dot{r}$ we use the normalization condition:
\begin{equation}\label{nor}
u^{\mu}u_{\mu}=1,\end{equation} here $u^{\mu}$ denotes the velocity
components of the moving particles. For the metric under
consideration, (\ref{nor}) yields
\begin{equation}\label{6}
f \dot{t}^2-\frac{\dot r ^2}{f}-r^2\dot\phi^2+2(\epsilon
r^2\Omega)\dot t\dot \phi=1.
\end{equation}
Making use of (\ref{4}) and (\ref{5}) in (\ref{6}), we get
\begin{equation}\label{7}
\dot{r}^2=f\Big(\frac{E^2}{m^2f}-\frac{L^2}{m^2r^2}-1 \Big),
\end{equation}
or
\begin{equation}\label{8}
\frac{dr}{d\tau}=\dot{r}=\pm \sqrt{f\Big(
\frac{E^2}{m^2f}-\frac{L^2}{m^2r^2}-1 \Big)},
\end{equation}
here $\pm$ sign denotes the radial velocity of the outgoing and
ingoing particles respectively.
\subsection{Geodesics of a Radially Moving Particle}
\begin{figure}
\includegraphics[width=10cm]{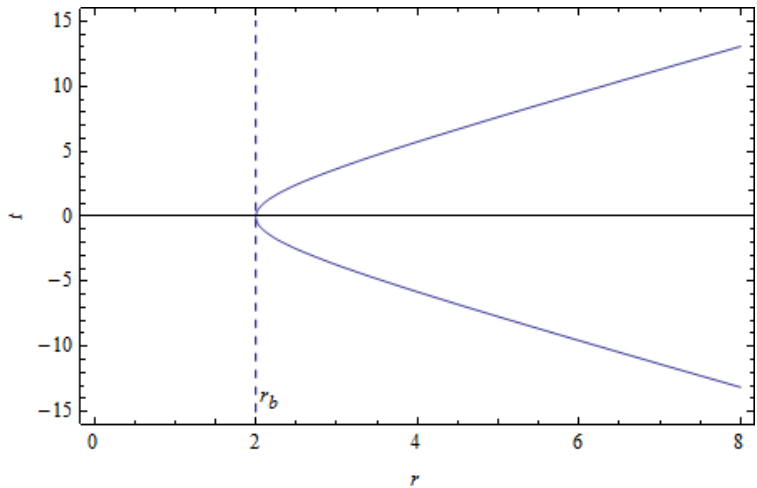}
\caption{Geodesics for a radially ingoing particle coming from
infinity with some initial velocity, reaching $r=r_b$ (dashed line),
and going back to infinity. We chose $\alpha=1$,
$\mathcal{E}^2=1.5$.} \label{BSW11}
\end{figure}

If the particle is moving radially towards black hole we can get the
geodesics of such particle. Using (\ref{5}) and (\ref{8}) together
for zero angular momentum, $L=0$, we obtain
\begin{equation}
\label{1a}\frac{d r}{d t}\equiv \frac{\dot{r}}{\dot{t}}
= \mp \frac{\sqrt{\mathcal{E}^2-f(r)}}{\mathcal{E}f^{-1}(r)},
\end{equation}
where $\mathcal{E}\equiv -E/ m$. Changing position of a radially
moving particle with the passage of time could be obtained from
(\ref{1a}), where positive root gives the path of the particle going
away from black hole, and negative root gives the path of an ingoing
particle. Note that (\ref{1a}) is
defined only if $\mathcal{E}^2> f(r)$, or
$r>\alpha/(\mathcal{E}^2-1)$. If we choose $\mathcal{E}^2=1.5$, then
there is a boundary, $r_b$,
\begin{equation} r_b \equiv r=\frac{\alpha}{0.5},\end{equation}
beyond which the particle can not go. Geodesics defined in
(\ref{1a}) could be understood better by plotting in $(r, t)$
coordinates, Fig(\ref{BSW11}).

\subsection{Circular Orbits and Behavior of Effective Potential}
\begin{figure}
\includegraphics[width=10cm]{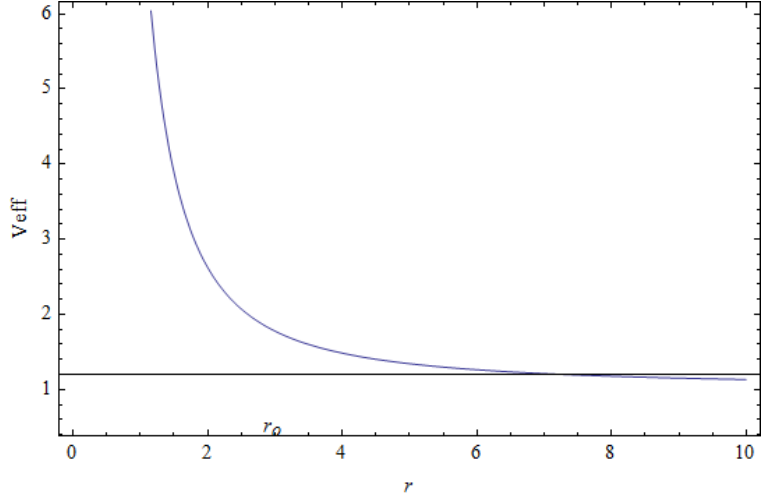}
\caption{Evolution of effective potential versus radius $r$,
parameters values are set as, $L^2=3$, $\alpha=1$, $m=1$.  At
$r=r_o=3$ there is a circular orbit of the particle. \label{Veff1}}
\end{figure}

Here we analyze the effective potential to look at the
stable and unstable circular orbits of the moving particles. It is
known that the possible range for the motion of the particle is
given by $E^2\geq V_\text{eff}^2$ (see for example \cite{EZ}).
Further the orbits of massive particles can be circular if the
$V_\text{eff}$ attains it maximum and minimum values. It means that
for the minimum value of $V_\text{eff}$ the circular orbits are
stable and for the maximum value of $V_\text{eff}$ the circular
orbits are unstable. For inflection point $V^{\prime \prime}=0$,
yields the marginally stable orbit,

We can write (\ref{7}) as below
\begin{equation}\label{Veff}
\dot{r}^2=\frac{1}{m^2 B^2}( E^2- V_\text{eff}),
\end{equation}
where
\begin{equation}
V_\text{eff}= f(\frac{L^2}{r^2}+m^2),
\end{equation}
is the effective potential and
\begin{equation}\label{deri}V^{'}_\text{eff}\equiv \frac{d V_\text{eff}}{dr}=\frac{-m^2 r^2 \alpha - L^2 (2 r + 3
\alpha)}{r^4}.
\end{equation} Solving $V^{'}_\text{eff}=0$, for $r$ we get
\begin{equation}\label{radii}
r_{\pm}=\frac{-L^2 \pm \sqrt{L^4-3L^2 m^2
\alpha^2}}{m^2 \alpha}.
\end{equation} For real values of $r$,
$L^2\geq 3 m^2\alpha^2$ is required.
Also
\begin{equation}
V^{\prime \prime }_\text{eff}\equiv \frac{d^2 V_\text{eff}}{dr^2}=
\frac{2(m^2 r^2 \alpha +3L^2 (r+2 \alpha))}{r^5}.
\end{equation}\\
Now we check the behavior of $V_\text{eff}$ at the above obtained
radii given by (\ref{radii}). Note that at $L^2 = 3 m^2 \alpha^2$,
particle has a circular orbit, of radius
$$|r_o|=3\alpha,$$ and $$V^{\prime
\prime}_\text{eff}\mid_{r=r_o}=0.$$ Behavior of $V_\text{eff} $ with
changing value of radius, $r$, is demonstrated in Fig.
(\ref{Veff1}).

If $L^2
> 3 m^2 \alpha^2$ then
\begin{equation}\label{positive}
V^{\prime \prime}_\text{eff}\mid_{r=r_+}= \frac{2 L^2 m^8 \alpha^4
(-L^2 + 3 m^2 \alpha^2 + \sqrt{ L^4 - 3 L^2 m^2 \alpha^2})}{(-L^2 +
\sqrt{ L^4 - 3 L^2 m^2 \alpha^2})^5},
\end{equation}
and
\begin{equation}\label{negative}
V^{\prime \prime}_\text{eff}\mid_{r=r_-}= \frac{2 L^2 m^8 \alpha^4
(L^2 - 3 m^2 \alpha^2 + \sqrt{ L^4 - 3 L^2 m^2 \alpha^2})}{(L^2 +
\sqrt{L^4 - 3 L^2 m^2 \alpha^2})^5}.
\end{equation}
Let $L^2-3\alpha^2 m^2=\beta^2$ the above expressions
(\ref{positive}) and (\ref{negative}) reduce to:
\begin{equation}\label{plus}
V^{\prime \prime}_\text{eff}\mid_{r=r_+}=\frac{2L^2 m^8 \alpha^4
(-\beta^2 + \beta L)}{(-L^2 +\beta L)^5},
\end{equation}\\
and
\begin{equation}\label{minus}
V^{\prime \prime}_\text{eff}\mid_{r=r_-}=\frac{2L^2 m^8 \alpha^4
(\beta^2 + \beta L)}{(L^2 +\beta L)^5},
\end{equation} We see that quantity in the R.H.S. of (\ref{plus}) is greater than zero if \\
 $$\beta L > \beta^2,\\~~ \beta L > L^2,$$ or if $$\beta^2>\beta L, ~~\\ L^2>L\beta,$$ but both conditions lead to the contradiction:
 \begin{equation}\label{con}
\beta>L>\beta.\end{equation}
Hence $V_\text{eff}$ does not attain minimum value at $r=r_+$. The
$V_\text{eff}$ has maximum at $r_+$ if $\beta<L$ (which gives
$3\alpha^2 m^2>0$, a physically true constraint) therefore the
circular orbits are unstable.

Now there are two possible cases for $V^{\prime \prime}_\text{eff}\mid_{r=r_-}>0$.\\ Case 1:\\
When both $\beta^2+\beta L$ and $L^2+\beta L$ are positive
quantities which give $\beta> -L$ or $3m^2\alpha^2<0$ which is a
contradiction so we exclude this case.
\\Case 2:\\
When both $\beta^2+\beta L< 0$ and $L^2+\beta L<0$. This implies
$\beta <-L$ or $3m^2 \alpha^2>0$, which is physically true. So
$V_\text{eff}\mid_{r=r_-}>0,$ if $\beta<-L$. Hence $V_\text{eff}$ is
minimum at $r=r_-$ if  $\beta<-L$ and the circular orbits are
stable.

\subsection{Center of Mass Energy}
Energy in the center of mass frame is defined as \cite{BSW}
\begin{equation}\label{ecm}
E_{cm}=m\sqrt{2}\sqrt{1-g_{\mu\nu}u^{\mu}_{1}u^{\nu}_{2}},
\end{equation}
where
\begin{equation}
u^{\mu}_i\equiv \frac{dx^{\mu}}{d\tau},~~~i=1, ~2
\end{equation}
is the $4$-velocity of each of the particle. Using (\ref{4}), (\ref{5}), and
(\ref{8}), ((\ref{8}) with negative sign as the particle coming
towards the black hole), in (\ref{ecm}), we get the CME, for the neutral
particle falling freely from rest at infinity as
\begin{eqnarray}\label{9}
E_{cm}&=& A \Big[ 1-\frac{E^2}{m^2f}+\frac{1}{m^2f}\Big\{E^4+
\frac{f}{r^2}
(L_1^2+L_2^2)(m^2f-E^2)\nonumber\\&&-fm^2(2E^2-fm^2)+\frac{L_1^2
L_2^2f^2}{r^4} \Big\}^{1/2} +\frac{L_1 L_2}{m^2r^2} \Big]^{1/2},
\end{eqnarray}
here $A=m\sqrt{2}$. We observe that near the horizon (i.e. at
$f(r)=0$), $E_{cm}$ in (\ref{9}) becomes undefined. However
expanding the numerator in (28), about the horizon yields
\begin{equation}\label{10}
E_{cm}=2m\sqrt{1+\frac{1}{4m^2 \alpha ^2}(L_1+ L_2)^2}.
\end{equation}
The CME in (\ref{10}) could be infinite, if the angular momentum of
one of the particles has infinite value, for which the particle
could not reach the horizon of the black hole. Thus the CME in
(\ref{10}) can not be unlimited.

\section{Motion of the Particles in the Vicinity of $\textbf{3+1}$ Dimensional Topological
Black Hole}
The metric of four dimensional Lifshitz black hole is
given by \cite{2009}
\begin{equation}\label{Lif}
ds^2=\frac{r^2 f(r)}{\ell^2}dt^2 -\frac{dr^2}{f(r)}-r^2 (d\theta^2
+\sinh^2\theta d\phi^2),
\end{equation}
where
\begin{equation}
f(r)=\frac{r^2}{\ell^2}-\frac{1}{2}.
\end{equation}
$f(r)=0$, gives the event horizon at $r_h=\ell/\sqrt{2}$, where
$\ell$ denotes the length scale in the geometry.
To find out the CME of the colliding particles in the vicinity of
Lifshitz black hole, we first find out the geodesics structure of
the particles. Using the standard Lagrangian procedure we find the
conserved quantities of the moving particles of mass $m$
($m=m_1=m_2$). For each particle the velocity components are given
by
\begin{equation} \label{E}
\dot{t}=-\sqrt{\frac{E}{m}}\frac{\ell^2}{fr^2},
\end{equation} and
\begin{equation}\label{L}
\dot{\phi}=\frac{L}{m r^2}.
\end{equation}
The normalization condition given in (\ref{nor}), used for the
metric defined in (\ref{Lif}) gives
\begin{equation}\label{r}
\dot{r}^2= \frac{\ell^2}{m r^2}(E-V_\text{eff}),
\end{equation}
or
\begin{equation}\label{x}
\dot{r}=\pm  \frac{\ell}{r\sqrt{m}}\sqrt{E-V_\text{eff}},
\end{equation}
the $\pm$ signs stand for radial velocities of outgoing and ingoing
particles respectively, and $V_\text{eff}$ is defined as
\begin{equation}\label{V}
V_\text{eff}=\frac{f r^2}{\ell^2}[m +\frac{L^2}{m r^2}].
\end{equation}
This expression for $V_\text{eff}$ is already derived in literature
\cite{topological}, we rederive it using dimensionally correct
equations (\ref{E}) and (\ref{L}). Geodesics of a radially moving
particle for this black hole are also already plotted in
\cite{topological}.

\subsection{Behavior of Effective Potential}
Differentiating (\ref{V}) with respect to $r$, we have
\begin{equation}
V^{'}_\text{eff}= \frac{r(2L^2 -m^2(\ell^2-4r^2))}{\ell^4 m}.
\end{equation}
Solving ${V^{\prime}}_\text{eff}=0$, for $r$ we get
\begin{equation}\label{radius}
r=0, ~~ r_{\pm}=\pm \frac{\sqrt{-2L^2+\ell^2 m^2}}{2m}.
\end{equation}
Furthermore,
\begin{equation}\label{effective}
 \frac{d^2 V_\text{eff}}{dr^2}\equiv V^{''}_\text{eff}=\frac{2L^2-m^2(\ell^2-12r^2)}{\ell^4 m}.
\end{equation}
We evaluate (\ref{effective}) at above obtained radii given in
(\ref{radius}), to check the nature of $V_\text{eff}$. At
$r=r_{\pm}$, (\ref{effective}) becomes
$$ \frac{d^2 V_\text{eff}}{dr^2}\equiv{V}^{\prime\prime}_\text{eff}= \frac{4L^2}{m \ell^4}+\frac{2m}{\ell^2}.$$ Note
that for stable orbit ${V}^{\prime\prime}_\text{eff}> 0$ then
 $$\frac{m^2 \ell^2}{2}>L^2.$$
It shows that $V_\text{eff}$ is minimum at $r=r_{\pm}$, for
$L^2<\ell^2/2$ and the two stable circular orbits exist. Choosing
$\ell=10$ as in \cite {topological}, we get
$-\sqrt{50}<L<\sqrt{50}.$ If the particles approach the black hole
from opposite directions then their angular momentums would have
opposite sign.

\subsection{Center of Mass Energy}
For two neutral particles falling freely from rest at infinity the
CME in the vicinity of $3+1$ dimensional topological Lifshitz black
hole is calculated by using (\ref{E}), (\ref{L}) and (\ref{x}) (in
(\ref{x}) we choose negative sign only, and its reason was explained
before in section 2) in equation of CME, we get
 \begin{equation}\label{topocme}E_{cm}=m\sqrt{2}\Big[(L_1^2 + L_2^2)(\frac{-1}{2r^2 m^2}
+\frac{f}{2 m E \ell^2 })+\frac{L_1 L_2}{m^2 r^2}(1+\frac{f L_1
L_2}{2 m E \ell^2})+\frac{f r^2 m}{2E \ell^2}\Big]^{1/2}.
\end{equation}
Expanding the numerator in (\ref{topocme}) at the horizon of the black we obtain
\begin{equation}\label{new}
E_{cm}=\frac{\sqrt{2}}{\ell}(L_1-L_2).
\end{equation}
Note that for the CME to be positive $L_1>L_2$.
The CME in (\ref{new}) goes to infinity if $L_1$
attain infinite value, for which the particle could not reach the
black hole horizon, hence in this case CME will remains finite.

\section{Conclusions}
In the background of a slowly rotating black hole in the HL theory of
gravity and $3+1$ dimensional topological Lifshitz black hole we
have analyzed the motion of massive (neutral) particles, falling
freely from infinity with zero velocity there. Analytical
expressions for effective potentials and the CME of neutral
particles have obtained by using the standard Lagrangian methods.
Geodesics of the particle coming from infinity, with some initial
velocity, approaching black hole radially, has also plotted in $(r,
t)$ coordinates. It has observed that in the case of slowly rotating
black hole in HL gravity the particle coming from infinity
approaches the black hole and a smooth curve has obtained which ends
at the boundary $r=r_b=2$. Then we have combined that curve with the
curve obtained for the particle moving away from
the black hole, Fig. (\ref{BSW11}) has obtained.

We have examined the behavior of effective potentials in both cases of
black holes. Calculations have shown that in the case of slowly rotating
black hole in the HL gravity, $V_\text{eff}$ in (\ref{positive}), does
not attain minimum value at $r=r_+$ given by (\ref{radii}) and is
maximum if $\beta<L$ is satisfied. It shows that the unstable
circular orbits of the particle exist there. While $V_\text{eff}$ in
(\ref{negative}) is minimum at $r=r_-$ if $\beta<-L$. In this case
there exists stable circular orbits of the moving particles. For the
$3+1$ dimensional topological Lifshitz black hole $V_\text{eff}$ in
(\ref{V}) is minimum at both $r=r_{\pm}$ given by (\ref{radius}),
for $-\sqrt{50}<L<\sqrt{50}$ and the circular orbits are stable. We
plot the effective potential of test particle moving around slowly
rotating black hole in HL gravity, as a function of $r$, by choosing
the appropriate values of parameters.

We also found the expressions for the collision energy in the CM frame
of neutral particles in the vicinity of these black holes. It is observed
that the CME of the neutral particles falling from rest at infinity will remain
finite in the CM frame at the horizon of these black holes. Thus the BSW
effect can not be seen for the two cases of the black holes discussed here.



\end{document}